\documentclass{article}
\usepackage{amssymb}

\newtheorem{proposition}{Proposition}
\begin{document}

\title{On some nondecaying potentials and related Jost solutions
for the heat conduction equation
\thanks{Work supported in part by PRIN 97 ``Sintesi''}}
\author{B. Prinari \thanks{e-mail: prinari@le.infn.it}\\
Dipartimento di Fisica dell'Universit\`{a} and Sezione INFN \\ 
I-73100 Lecce \\ ITALY}
\maketitle

\begin{abstract}
Potentials of the heat conduction operator constructed by means of $2$ binary
B\"{a}cklund transformations are studied in detail. Corresponding Darboux
transformations of the Jost solutions are introduced. We show that these
solutions obey modified integral equations and present their analyticity
properties.
\end{abstract}

\section{Introduction}

In this article we investigate into the direct and inverse scattering
transform for the heat conduction operator 
\begin{equation}
L=-\partial _{x_{2}}+\partial _{x_{1}}^{2}-u(x_{1},x_{2}),  \label{1}
\end{equation}
in the case in which $u$ is a real function with ``ray'' type behavior. More
exactly, $u$ is supposed to be rapidly decaying in all directions on the 
$x$-plane with the exception of some finite number of directions, where it has
finite and nontrivial limits, i.e. \begin{equation}
u_{n,\pm }(x_{1})=\lim_{x_{2}\rightarrow \pm \infty }u(x_{1}-2\mu
_{n}x_{2},x_{2}),\qquad n=1,2,\dots ,N,  \label{u+-}
\end{equation}
for $N$ real constants $\mu _{n}$. The spectral theory of operator $L$ with
potential in this class is interesting \textit{per se} and because it is
associated to the Kadomtsev--Petviashvili equation~in its version called
KPII \cite{Kadomtsev-P}--\cite{zakharov74} 
\begin{equation}
(u_{t}-6uu_{x_{1}}+u_{x_{1}x_{1}x_{1}})_{x_{1}}=-3u_{x_{2}x_{2}}.
\label{KPII}
\end{equation}

In investigating KPII equation it was found \cite{ABYF} that there are
significant differences with respect to the case of KPI (equation (\ref{KPII})
with opposite sign in the rhs). Specifically, the
inverse problem cannot be formulated as a Riemann--Hilbert boundary value
problem \cite{ZManakov}--\cite{FokasAblowitz}. 
This results from the fact that for the associated spectral problem
there exist eigenfunctions which, though bounded, are nowhere analytic in
the spectral parameter. The role of the Riemann--Hilbert problem is now
played by what is usually called a $\bar{\partial}$-problem. The basic idea
is to compute the $\bar{\partial}$-derivative of these eigenfunctions
(modified Jost solutions) and then to exploit the Cauchy--Green formula to
get a linear Fredholm integral equation for such solutions (under the
assumption that the homogeneous integral equation has no nontrivial
solutions). One can prove the smoothness of the potential $u(x,t)$ and also
some other properties \cite{Grinevich}. Anyway, it is always assumed that
the potential is rapidly decaying for $x_{1}^{2}+x_{2}^{2}\rightarrow \infty 
$ (which excludes in particular the presence of line solitons) and a
rigorous investigation of the IST and of the properties of solutions of KPII
equation with nontrivial (for instance, one-dimensional) asymptotic behavior
is still missing. Such extension of the spectral theory for the heat operator
would provide the possibility to extend correspondingly the class of
solutions of the KPII equation.

The same problem has been faced for the nonstationary Schr\"odinger equation 
in the framework of the
so-called resolvent approach \cite{nsreshi}--\cite{towards}. This extension
in the general case resulted to be particularly involved, in particular as
far as the definition of spectral data and their characterization equations
are concerned. 
Some general results were obtained in \cite{KPlett} with some
(partially implicit) assumptions. Even though it was not possible to define
precisely the subclass of potentials satisfing these assumptions, it was
shown that potentials belonging to this special subclass can be obtained via
B\"{a}cklund transformations (BT's). For getting some experience with a
case less involved from a mathematical point of view but sufficiently
general, it was considered the case of potentials obtained by applying
recursively binary BT's to an arbitrary decaying potential and describing $N$
solitons superimposed to a generic background \cite{steklov}. At least in
the case $N=1$ it was possible to obtain explicitly the resolvent, the Jost
solutions and the spectral data and to show that they satisfy all the
required assumptions \cite{towards}.  Moreover, there is evidence that the 
structural properties of the spectral data can be even more complicated in 
the general case.

Spectral theory of the operator~(\ref{1}) with ray potential is essentially
more involved than the standard case of rapidly decaying potential. One of
the problems is that the standard integral equations defining the Jost
solution and its dual become senseless, since the Green's function is slowly
decaying at space infinity and cannot ensure convergency. 
In Sec.~2, after reviewing some properties of Green's function and
Jost solutions for the heat conduction equation in the case of 
rapidly decaying potentials, we propose an
extension of these integral equations which allows also to include
potentials with ray type behavior and which in the one-dimensional limit
give the standard integral equations for the Jost solutions of KdV equation.
We call the solutions of these modified integral equations Jost-like solutions.
In Sec.~3 we consider the potentials obtained by superimposing, via binary
B\"{a}cklund transformations, one soliton to a generic decaying background.
A significant difference with respect to the case of the nonstationary 
Schr\"{o}dinger operator is due to the fact that in order to get a real
potential we are obliged to use as functions generating the BT's the Jost 
solution and dual Jost solution computed for purely imaginary values of the 
spectral parameter and thus an accurate study of their properties at these 
values is required. We get the explicit
expression of the potential and we are able to formulate the conditions to
be imposed on the parameters of the B\"{a}cklund transformation in order to
obtain a real and regular solution. Moreover, we obtain also the
explicit expression of the Jost-like solutions and study their spectral 
properties.
Finally, in Sec.~4 we iterate this procedure to get two solitons on a generic
background. Also in this case we determine the regularity conditions for the
solutions obtained by these means and we explicitly compute Jost solutions
and study their spectral properties. An interesting feature of these
solutions describing two solitons on a background is that, unlike KPI case,
the directions of the solitons are not fixed by the parameters of the
B\"{a}cklund transformations but depend also on the ``constants'' of
integration of the Darboux procedure. If the matrix which appears
in the expression of the potential is diagonal, the two BT's simply
``superimpose'' the two solitons with only a phase shift. But in the general
case, that is when both off-diagonal entries of such matrix are different
from zero, the solitons are rotated with respect to the directions
determined by the single BT's. When one of the off-diagonal entries is null
but the other one is not, only one soliton is rotated while the other one is
unchanged and we could say that the ``interaction'' does not preserve the
soliton direction. If a row or a column in matrix $C_{2}$ is made up of zeros
the resulting potential has three ``rays''.
Note that all these features are absent in the case of KPI
equation. In fact, for KPI the soliton directions do not depend on the
choice of such matrix \cite{steklov}. Another difference with respect to the
case of KPI equation is that the KPI\ wave soliton emerging from the
background at large distances divides the $x$-plane into two regions with
different asymptotic behavior. In the first region (to the left of the
soliton) the original potential $u(x)$ is modified at large distances by an
exponentially decreasing term, while in the second region (to the right) by
a term decreasing as $1/x_{2}^{3}$. In the case of KPII the corrections are
always decaying at infinity faster that any power of $x_{2}$.

\section{On Jost solutions}

In the standard case one introduces the complex spectral parameter 
$k=k_{\Re}+ik_{\Im}$ and defines the Jost solution $\Phi (x,k)$ and dual
Jost solution $\Psi (x,k)$ as the solutions of the heat conduction equation
and its dual, respectively, such that 
\begin{eqnarray}
\chi (x,k) &=&e^{ikx_{1}+k^{2}x_{2}}\Phi (x,k)  \label{3} \\
\xi (x,k) &=&e^{-ikx_{1}-k^{2}x_{2}}\Psi (x,k)  \label{3dual}
\end{eqnarray}
satisfy the integral equations 
\begin{eqnarray}
\chi (x,k) &=&1+\int dx^{\prime }G_{0}(x-x^{\prime },k)u(x^{\prime })\chi
(x^{\prime },k)  \label{4} \\
\xi (x,k) &=&1+\int dx^{\prime }G_{0}(x^{\prime }-x,k)u(x^{\prime })\xi
(x^{\prime },k)  \label{4dual}
\end{eqnarray}
where the Green's function  $G_{0}(x,k)$ is given by 
\begin{equation}
G_{0}(x,k)=-\frac{\mathop{\rm sign}\nolimits x_{2}}{2\pi }\int d\alpha
\theta (\alpha (\alpha +2k_{\Re })x_{2})e^{-i\alpha x_{1}-\alpha (\alpha
+2k)x_{2}}.  \label{G0}
\end{equation}

It is evident from expression (\ref{G0}) that $G_{0}$ and, consequently, 
$\chi $ and $\xi $ are nowhere analytic in the complex $k$-plane. Functions 
$\chi $ and $\xi $ are bounded for all $x_{1},x_{2}$ (see for instance 
\cite{Clarksonbook}) and 
\[
\lim_{k\rightarrow \infty }\chi (x,k)=\lim_{k\rightarrow \infty }\xi (x,k)=1.
\]
The Jost solutions defined by (\ref{3}) and (\ref{3dual}) obey the following
normalization and completness conditions 
\begin{eqnarray}
&&\int dx_{1}\Psi (x_{1},x_{2},k+p)\Phi (x_{1},x_{2},k)=2\pi \delta (p)
\label{xortog} \\
&&\int dk_{\Re }\Psi (x_{1}^{\prime },x_{2},k)\Phi (x_{1},x_{2},k)=2\pi
\delta (x_{1}-x_{1}^{\prime })  \label{xcompl}
\end{eqnarray}

One can check that 
\begin{equation}
G_{0}(x,k)=\frac{\mathop{\rm sign} k_{\Re }}{2i\pi \left| x_{1}\right| }
\left[e^{2ik_{\Re }(x_{1}+2k_{\Im }x_{2})}-1\right] +o(\left|
x_{1}\right|^{-1})\qquad  \label{asympx1}
\end{equation}
and 
\begin{equation}
G_{0}(x,k)=-\frac{\mathop{\rm sign}k_{\Re }}{4\pi k\ x_{2}} +o(\left|
x_{2}\right|^{-1}).  \label{asympx2}
\end{equation}
Despite these asymptotics are discountinuous at $k_{\Re }=0$, the Green's
function is not and one can easily check from (\ref{G0}) that 
\[
G_{0}(x,ik_{\Im })=-\frac{\theta (x_{2})}{2\sqrt{\pi x_{2}}}
e^{-\frac{(x_{1}+2k_{\Im }x_{2})^{2}}{4x_{2}}} 
\]
which is real and, up to the exponential factor $e^{-k_{\Im }x_{1}-k_{\Im
}^{2}x_{2}},$ is the well-known Gauss-Weierstrass kernel (or heat kernel).
Then $\chi $ and $\xi $ can be computed at purely imaginary values of the
spectral parameter $k$ and they are real as well.

Solvability of the integral equations (\ref{4})--(\ref{4dual})
under some small norm assumptions was
proved in~\cite{Wick} and thanks to (\ref{asympx1}) it is easy to show that 
$\chi (x,k)$ and $\xi (x,k)$ have the asymptotic behavior 
\begin{eqnarray}
\lim_{\left| x_{1}\right| \rightarrow \infty }\chi (x,k) &=&1  \label{chix1}
\\
\lim_{\left| x_{1}\right| \rightarrow \infty }\xi (x,k) &=&1  \label{xix1}
\end{eqnarray}
Moreover, one can prove the following propositions.

\begin{proposition}
Let $u$ be a real, regular and rapidly decaying potential. The modified
Jost solutions $\chi$ and $\xi$ obey for any $n\in \mathbb{N}$ the
following asymptotics 
\begin{eqnarray}
\lim_{x_{2}\rightarrow \pm \infty }x_{2}^{-n}\left( \chi (x,ik_{\Im}
)-1\right)  &=&0,  \label{asympchi} \\
\lim_{x_{2}\rightarrow \pm \infty }x_{2}^{-n}\left( \xi (x,ik_{\Im}
)-1\right)  &=&0.  \label{asympxi}
\end{eqnarray}
\end{proposition}
{\it Proof}. 
Let us consider the integral equation (\ref{4}) for $\chi$.
We have for $k_{\Im}=\kappa \in \mathbb{R}$
\begin{equation}
\chi (x,i\kappa )-1=-\frac{1}{2\sqrt{\pi }}\int dx^{\prime }\frac{\theta
(x_{2}-x_{2}^{\prime })}{\sqrt{x_{2}-x_{2}^{\prime }}}e^{-\frac{\left(
x_{1}-x_{1}^{\prime }+2\kappa (x_{2}-x_{2}^{\prime })\right) ^{2}}
{4(x_{2}-x_{2}^{\prime })}}u(x^{\prime })\chi (x^{\prime },i\kappa ) 
\nonumber
\end{equation}
and we can make shift and proper change of variables to get 
\begin{equation}
\chi (x,i\kappa )-1=-\frac{1}{\sqrt{\pi }}\int dx^{\prime }\theta
(x_{2}^{\prime })e^{-(x_{1}^{\prime })^{2}}(u\chi )(x_{1}-2x_{1}^{\prime }
\left| x_{2}^{\prime }\right|^{1/2} +2\kappa \left| x_{2}^{\prime }\right|
,x_{2}-\left| x_{2}^{\prime }\right| ,i\kappa ).  \nonumber
\end{equation}
Note that 
\begin{eqnarray*}
&&\left| \int dx_{1}^{\prime }\int_{0}^{+\infty }dx_{2}^{\prime }\
e^{-(x_{1}^{\prime })^{2}}(u\chi )(x_{1}-2x_{1}^{\prime }\left|
x_{2}^{\prime }\right| ^{1/2}+2\kappa \left| x_{2}^{\prime }\right| ,x_{2}-
\left|x_{2}^{\prime }\right| ,i\kappa )\right| \le  \\
&\le &\int_{-\infty }^{+\infty }dx^{\prime }\ e^{-(x_{1}^{\prime
})^{2}}\left| (u\chi )(x_{1}-2x_{1}^{\prime }\left| x_{2}^{\prime
}\right| ^{1/2} +2\kappa \left| x_{2}^{\prime }\right| ,x_{2}-\left| 
x_{2}^{\prime}\right| ,i\kappa )\right| .
\end{eqnarray*}
Since $\chi $ is bounded and $u$ is in the Schwartz class 
the limit $\left| x_{2}\right| \rightarrow \infty $ can be exchanged with
the integral and the integrand decays as fast as $u$, that is faster than
any power of $x_{2}$. The same result holds for $\xi $.

\begin{proposition}
Let $u$ be a real potential in the Schwartz class. The modified Jost
solutions $\chi $ and $\xi $ for the  heat conduction equation
with potential $u $ are such that $\chi (x,ik_{\Im })\neq 0,\ \xi
(x,ik_{\Im })\ne 0$ $\forall x\in \mathbb{R}^{2},$ or equivalently, taking
into account asymptotics (\ref{asympchi})--(\ref{asympxi}), 
\begin{eqnarray}
\chi (x,ik_{\Im }) &>&0\qquad   \label{positiv} \\
\xi (x,ik_{\Im }) &>&0,\qquad \forall x\in \mathbb{R}^{2}.
\end{eqnarray}
\end{proposition}
{\it Proof}. In order to simplify the notation in the following we will
put $k_{\Im }=\kappa \in \mathbb{R}$. Let us write the
solution of the integral equation (\ref{4}) as the (formal) Neumann
series 
\[
\chi (x,i\kappa )=\sum_{n=0}^{\infty }\chi _{n}(x,i\kappa )
\]
where 
\begin{eqnarray*}
\chi _{0}(x,i\kappa ) &=&1 \\
\chi _{n}(x,i\kappa ) &=&-\frac{1}{2\sqrt{\pi }}\int dx^{\prime}
\frac{\theta (x_{2}-x_{2}^{\prime })}{\sqrt{x_{2}-x_{2}^{\prime }}}
e^{-\frac{\left[ \left( x_{1}-x_{1}^{\prime }\right) +2\kappa 
(x_{2}-x_{2}^{\prime})\right] ^{2}}{4(x_{2}-x_{2}^{\prime })}}u 
\left( x^{\prime }\right)\chi _{n-1}(x^{\prime },i\kappa )
\end{eqnarray*}
Obviously we have to consider the class of potentials for which the
existence of solution of the integral equation, that is the convergency of
the Neumann series, is estabilished (see \cite{Wick} and the following
remark). \newline
Now let us assume that for all $x \in \mathbb{R}^{2}$
\[
\left| u(x)\right| <U(x_{2}) \label{boundu}
\]
with $U$ such that
\[
\int_{0}^{+\infty }dx_{2}^{\prime }\left| U(x_{2}-x_{2}^{\prime })\right| <1.
\label{boundU}
\]
If we introduce
\begin{equation}
a(x)=\int dx^{\prime }\frac{\theta (x_{2}-x_{2}^{\prime })}
{\sqrt{x_{2}-x_{2}^{\prime }}}e^{-\frac{\left[ \left( x_{1}-x_{1}^{\prime }
\right)+2\kappa (x_{2}-x_{2}^{\prime })\right] ^{2}}
{4(x_{2}-x_{2}^{\prime })}}\left| u(x^{\prime })\right| ,  \label{def:a}
\end{equation}
due to (\ref{boundu})--(\ref{def:a}) we have
\begin{equation}
0<\frac{a(x)}{2\sqrt{\pi}}<\frac{1}{2}.  \label{a0}
\end{equation}
Then by induction on $n$ one can easily prove that 
\[
\left| \chi _{n+1}(x,i\kappa )\right| <\left( \frac{1}{2}\right) ^{n+1},
\]
so that 
\[
\left| \chi (x,i\kappa )-1\right| <1.
\]

The main problem when dealing with a potential $u(x)$ not vanishing in all
directions at large distances is that the integral equations~(\ref{4}) and 
(\ref{4dual}) cannot be applied as the Green's function is slowly decaying at
space infinity. We suggest the following modification of these integral
equations: 
\begin{eqnarray}
\chi (x,k) &=&1+\hspace{-2pt}\int\limits_{-k_{\Im }\infty }^{\qquad x_{1}}
\hspace{-6pt}dy_{1}\int dx^{\prime }\partial
_{y_{1}}G_{0}(y_{1}-x_{1}^{\prime },x_{2}-x_{2}^{\prime },k)u(x^{\prime
})\chi (x^{\prime },k),  \label{12} \\
\xi (x,k) &=&1+\hspace{-2pt}\int\limits_{k_{\Im }\infty }^{\qquad x_{1}}
\hspace{-6pt}dy_{1}\int dx^{\prime }\partial _{y_{1}}G_{0}(x_{1}^{\prime
}-y_{1},x_{2}^{\prime }-x_{2},k)u(x^{\prime })\xi (x^{\prime },k),
\label{13}
\end{eqnarray}
where the order of operations is explicitly prescribed. Here and below we
use notations of the type $k_{\Im }\infty $ in the limits of integrals to
indicate the sign of infinity. If the solution of this equation exists and
is bounded, then like in the standard case 
\begin{eqnarray}
\lim_{x_{1}\rightarrow -k_{\Im }\infty }\chi (x,k) &=&1,\qquad  \label{asx}
\\
\lim_{x_{1}\rightarrow k_{\Im }\infty }\xi (x,k) &=&1,\qquad k_{\Im }\neq 0,
\end{eqnarray}
while in contrast to~(\ref{chix1}) and (\ref{xix1}) they can be different
from 1 in the opposite direction. These modified integral equations are
applicable to the simplest case of a potential of type~(\ref{u+-}), i.e.\ to
the case $u(x)=u(x_{1})$ and one recovers the standard one-dimensional
integral equation for Jost solution of KdV equation. 

Since the general case is
extremely involved, we are studying here the special but rather wide
subclass of potentials of type~(\ref{u+-}) that is obtained by applying
recursively the so-called binary B\"{a}cklund transformations~\cite{matveev}
with complex spectral parameter to a decaying potential. As we are
interested in the spectral properties of potentials $u$ having
nontrivial limits, following an approach close to the one in \cite{darboux}, 
we study also the corresponding 
Darboux transformations furnishing the Jost solutions of the transformed 
potentials and their analytical properties, as well as transformations of the 
spectral data.

\section{Binary B\"{a}cklund transformations}

We have at our disposal in~\cite{matveev} and~\cite{salle} a rather simple
and transparent method for performing binary B\"{a}cklund transformations of
the potential $u$ and corresponding Darboux transformations of solutions of~
(\ref{1}) and its dual. Let $u_{0}$ be a smooth, real and rapidly decaying
potential and $L_{0}$ the corresponding heat operator

\begin{equation}
L_{0}=-\partial _{x_{2}}+\partial _{x_{1}}^{2}-u_{0}(x).  \label{heatop}
\end{equation}
A new potential $u_{0}^{\prime }(x)$ can be generated through an elementary
BT 
\begin{equation}
L_{0}^{\prime }B_{0}=B_{0}L_{0}  \label{BT}
\end{equation}
using a gauge operator 
\begin{equation}
B_{0}=\partial _{x_{1}}-\left( \partial _{x_{1}}\log \varphi _{0}(x)\right) .
\label{gaugeB}
\end{equation}
If we impose (\ref{BT}) we find that 
\begin{equation}
u_{0}^{\prime }(x)=u_{0}(x)-2\partial _{x_{1}}^{2}\log \varphi _{0}(x)
\label{u'0}
\end{equation}
and that $\varphi _{0}$ has to be a solution of the original spectral
problem, that is 
\begin{equation}
\left( -\partial _{x_{2}}+\partial _{x_{1}}^{2}-u_{0}(x)\right) \varphi
_{0}(x)=0.  \label{phi}
\end{equation}

Let us consider also the spectral operator dual to (\ref{heatop}), that is 
\begin{equation}
L_{0}^{d}=\partial _{x_{2}}+\partial _{x_{1}^{2}}-u_{0}(x)  \label{dualheat}
\end{equation}
and the gauge operator 
\begin{equation}
B_{0}^{d}=\partial _{x_{1}}+\left( \partial _{x_{1}}\log \varphi
_{0}(x)\right) .  \label{dualgauge}
\end{equation}
Eq. (\ref{BT}) is equivalent to the corresponding dual relation 
\begin{equation}
B_{0}^{d}L_{0}^{\prime \ d}=L_{0}^{d}B_{0}^{d}.  \label{dualBT}
\end{equation}
Moreover, given any two solutions of the spectral problem (\ref{heatop}) and 
of its dual (\ref{dualheat}), say $\phi_{0}$ and $\psi_{0}$, their Darboux
transforms $\phi^{\prime}_{0}$ and $\psi^{\prime}_{0}$, defined by
\begin{eqnarray}
\phi _{0}^{\prime }=B_{0}\phi _{0}  \label{111} \\
B_{0}^{d}\psi _{0}^{\prime }=\psi _{0}  \label{113}
\end{eqnarray}
solve, respectively,
the spectral problem for the transformed potential and its dual up to
annihilators of $B_{0}^{d}$, that is
$L_{0}^{\prime }\phi _{0}^{\prime }=0$ and 
$B_{0}^{d}L_{0}^{\prime }\psi _{0}^{\prime }=0$.

One can easily check that if $f$ and $g$ solve (\ref{phi}) and its dual,
respectively, then 
\begin{equation}
\partial _{x_{2}}\left( f(x)g(x)\right) =-\partial _{x_{1}}W(f(x),g(x))
\label{partx2}
\end{equation}
where $W$ is standard Wronskian of $f$ and $g$, that is
\begin{equation}
W(f(x),g(x))=f(x)\partial _{x_{1}}g(x)-g(x)\partial _{x_{1}}f(x).
\label{Wronsk}
\end{equation}

Now let us consider an inverse elementary BT 
\begin{equation}
L_{0}^{\prime }B_{0}^{\prime }=B_{0}^{\prime }L_{1}  \label{120}
\end{equation}
through the gauge 
\begin{equation}
B_{0}^{\prime }=\partial _{x_{1}}+\left( \partial _{x_{1}}\log \psi
_{0}^{\prime }(x)\right) .  \label{126}
\end{equation}
In this case we have 
\begin{equation}
u_{1}(x)=u_{0}^{\prime }(x)-2\partial _{x_{1}}^{2}\log \psi _{0}^{\prime }
\label{125}
\end{equation}
where $\psi _{0}^{\prime }$ is an arbitrary solution of 
\begin{equation}
\left( \partial _{x_{2}}+\partial _{x_{1}}^{2}-u_{0}^{\prime }(x)\right)
\psi _{0}^{\prime }(x)=0.  \label{124}
\end{equation}
Now we perform direct and inverse elementary B\"{a}cklund transformations 
(\ref{BT}) and (\ref{dualBT}) and study the properties of operator $L_{1}$,
omitting all intermediate constructions associated to operator 
$L_{0}^{\prime }$. First of all, from (\ref{u'0}) and (\ref{125}) we see that 
\begin{equation}
u_{1}(x)=u_{0}(x)-2\partial _{x_{1}}^{2}\log \Delta _{1}(x)  \label{127}
\end{equation}
where we introduced 
\begin{equation}
\Delta _{1}(x)=\varphi _{0}(x)\psi _{0}^{\prime }(x).  \label{128}
\end{equation}
Moreover, using (\ref{u'0}), from (\ref{phi}) and (\ref{124}) we find that 
$\varphi _{0}$ and $\psi _{0}^{\prime }$ have to be chosen in such a way that
they satisfy the following system of equations 
\begin{equation}
\left\{ 
\begin{array}{l}
(-\partial _{x_{2}}+\partial _{x_{1}}^{2}-u_{0}(x))\varphi _{0}(x)=0 \\ 
\left( \partial _{x_{2}}+\partial _{x_{1}}^{2}-2\left( \partial _{x_{1}}\log
\varphi _{0}(x)\right) \partial _{x_{1}}\right) \Delta _{1}(x)=0
\end{array}
\right. .
\end{equation}
It is clear that a sufficient condition for $u_{1}$ in (\ref{127}) to be
real is to choose $\Delta _{1}$ real, but this means that $\varphi _{0}$
cannot be an arbitrary solution of the original spectral problem. In
particular, one can choose a real solution such as the Jost solution
computed for purely imaginary value of spectral parameter, that is 
\begin{equation}
\varphi _{0}(x)=\Phi _{0}(x,i\kappa _{1}).  \label{132}
\end{equation}
In order to obtain $\psi _{0}^{\prime }(x)$, and consequently $\Delta _{1}$, we
consider the Darboux transform (\ref{113}) of a solution $\psi _{0}(x)$ 
of the spectral problem $L_{0}^{d}\psi _{0}=0$ and we choose 
\begin{equation}
\psi _{0}(x)=\Psi _{0}(x,i\alpha _{1})  \label{psi}
\end{equation}
where $\Psi_{0}$ is the Jost solution of the spectral problem 
(\ref{dualheat}) computed for $k=i\alpha _{1},$ $\alpha_{1} \in \mathbb{R}$.
With this choice we get
\begin{equation}
\psi _{0}^{\prime }(x)=\frac{1}{\varphi _{0}(x)}\left[
C(x_{2})+\int_{-(\kappa _{1}-\alpha _{1})\infty }^{x_{1}}dx_{1}^{\prime
}\Psi _{0}(x_{1}^{\prime },x_{2},i\alpha _{1})\Phi _{0}(x_{1}^{\prime
},x_{2},i\kappa _{1})\right]  \label{136}
\end{equation}
and the integral is convergent due to the asymptotic behaviors (\ref{chix1}) 
and (\ref{xix1}). ``Constant'' of integration $C$ depends, in general, on 
$x_{2}$ as well, but one can check that $\psi_{0}^{\prime }$ is indeed a 
solution of (\ref{124}) iff $C$ does not depend on $x_{2}$ . So finally we get
\begin{equation}
\Delta _{1}(x)=c_{1}+\int_{-(\kappa _{1}-\alpha _{1})\infty
}^{x_{1}}dx_{1}^{\prime }\Phi _{0}(x_{1}^{\prime },x_{2},i\kappa _{1})\Psi
_{0}(x_{1}^{\prime },x_{2},i\alpha _{1}),  \label{137}
\end{equation}
where $c_{1}$ and $\alpha _{1},\kappa _{1}$ are real constants with $\alpha
_{1}\ne \kappa _{1}$.

Taking into account (\ref{137}), it is always possible
to formulate a small norm condition which ensures regularity of the dressed
potential (\ref{127}). Indeed, Prop.~2 proves that $\Phi_{0}(x,ik_{\Im })$ and 
$\Psi _{0}(x,ik_{\Im })$ are real and strictly
positive for all $x\in \mathbb{R}^{2}$; then the integral
in (\ref{137}) is a monotonic function of $x_{1}$ and if we choose $c_{1}$
such that 
\begin{equation}
(\kappa _{1}-\alpha _{1})c_{1} \ge 0,  \label{138}
\end{equation}
eq.~(\ref{137}) gives a function $\Delta _{1}$ which is regular, has no
zeros in the $x$-plane and has the same sign as $\kappa _{1}-\alpha _{1}$.
Consequently, condition (\ref{138}) ensures regularity of potential (\ref
{127}). Finally, one can easily show that $\Delta _{1}$ indeed superimposes
one soliton on the generic (smooth and rapidly decaying) background $u_{0}$
along the direction 
$x_{1}+(\kappa _{1}+\alpha_{1})x_{2} ={\mathrm const}$.

\subsection{Darboux procedure}

We need now to express the Jost solutions $\Phi _{1}$ and $\Psi _{1}$ of the
spectral equations 
\begin{equation}
L_{1}\Phi _{1}=0, \qquad  L_{1}^{d}\Psi _{1}=0  \label{210}
\end{equation}
in terms of the Jost solutions $\Phi _{0}$ and $\Psi _{0}$ of the original
spectral problem, that is we have to construct the Darboux version of the
binary BT. From (\ref{120}) we see that if 
$\phi _{0}^{\prime }$ satisfies $L_{0}^{\prime }\phi _{0}^{\prime }=0$ 
and we define $\phi _{1}$ through 
\begin{equation}
B_{0}^{\prime }\phi _{1}=\phi _{0}^{\prime },  \label{213}
\end{equation}
then $\phi _{1}$ is such that $B_{0}^{\prime }L_{1}\phi _{1}=0$.
So, taking into account definition of $B_{0}^{\prime }$ in (\ref{126}), from
(\ref{213}) we get 
\begin{equation}
\partial _{x_{1}}\left( \phi _{1}(x,k)\psi _{0}^{\prime }(x)\right) =\phi
_{0}^{\prime }(x,k)\psi _{0}^{\prime }(x).  \label{ident2}
\end{equation}
In order to have transformations parametrized by constants and not by
functions of $x_{2}$ obeying some differential equation, it would be 
convenient to integrate this equation from infinity to $x_{1}.$ 
From the other side, it is natural to use (\ref{111}) to get 
$\phi_{0}^{\prime}$, that is 
\begin{equation}
\phi _{0}^{\prime }(x,k)=\partial _{x_{1}}\phi _{0}(x,k)-\frac{\partial
_{x_{1}}\Phi _{0}(x,i\kappa _{1})}{\Phi _{0}(x,i\kappa _{1})}\phi _{0}(x,k),
\label{Bphi}
\end{equation}
and in particular to choose $\phi _{0}(x,k)\equiv \Phi _{0}(x,k)$. 
However, in this case $\phi _{0}^{\prime }$
has at large $x_{1}$ the same behavior as $\Phi _{0}$, and then from 
(\ref{136}) it follows that for $(k_{\Im }-\alpha _{1})(k_{\Im}-\kappa_{1})<0$
it is not possible to integrate (\ref{ident2}) on the infinite interval. 
Then we integrate (\ref{ident2}) from some $x_{1}^{0}$ to $x_{1}$ getting 
\begin{equation}
\phi _{1}(x,k)=\frac{1}{\psi _{0}^{\prime }(x)}\left[ C_{1}^{\prime
}(x_{2},k)+\int_{x_{1}^{0}}^{x_{1}}dx_{1}^{\prime }\phi _{0}^{\prime
}(x_{1}^{\prime },x_{2},k)\psi _{0}^{\prime }(x_{1}^{\prime },x_{2})\right]. 
\label{phitilde}
\end{equation}   
$C_{1}^{\prime }(x_{2},k)$ is a ``constant'' of integration and $\phi
_{1}(x,k)$ is a solution of $L_{1}\phi _{1}=0$ iff 
\[
\partial _{x_{2}}\left( C_{1}^{\prime }(x_{2},k)\right) =-\left. W\left(
\phi _{0}^{\prime }(x,k),\psi _{0}^{\prime }(x)\right) \right|
_{x_{1}=x_{1}^{0}}. \label{Dx2C'}
\]
Substituting $\phi_{0}^{\prime }$ given by (\ref{Bphi}) into (\ref{phitilde})
and integrating by parts we get 
\begin{equation}
\phi _{1}(x,k)=\Phi _{0}(x,k)+\frac{1}{\psi _{0}^{\prime }(x)}\left[
C_{1}(x_{2},k)-\int_{-(k_{\Im }-\alpha _{1})\infty}^{x_{1}}
\hspace{-20pt}dx_{1}^{\prime }\ \Phi _{0}(x_{1}^{\prime },x_{2},k)\Psi
_{0}(x_{1}^{\prime },x_{2},i\alpha _{1})\right] ,  \label{finite}
\end{equation}
where 
\begin{equation}
C_{1}(x_{2},k)=C_{1}^{\prime }(x_{2},k)-\Phi _{0}(x_{1}^{0},x_{2},k)\psi
_{0}^{\prime }(x_{1}^{0},x_{2})
+\int_{-(k_{\Im }-\alpha _{1})\infty }^{x_{1}^{0}}\hspace{-20pt}
dx_{1}^{\prime}\ \Phi _{0}(x_{1}^{\prime },x_{2},k)\Psi _{0}(x_{1}^{\prime },
x_{2},i\alpha_{1}). \label{C}
\end{equation}
One can check that due to (\ref{Dx2C'}) and (\ref{C}) 
$C_{1}$ does not depend on $x_{2}$. 
Finally,  we obtain from (\ref{phitilde}) a solution 
of the heat conduction equation with potential $u_{1}$ 
parametrized by an arbitrary function of $k$ 
\begin{equation}
\phi _{1}(x,k)=\Phi _{0}(x,k)-\frac{\Phi _{0}(x,i\kappa _{1})}
{\Delta _{1}(x)}\left[ C_{1}(k)+\int_{-(k_{\Im }-\alpha _{1})\infty }^{x_{1}}
\hspace{-20pt}dx_{1}^{\prime}\ \Phi _{0}(x_{1}^{\prime },x_{2},k)
\Psi _{0}(x_{1}^{\prime },x_{2},i\alpha_{1})\right]   \label{217}
\end{equation}
and we can introduce also the solution $F_{1}$ defined by 
\begin{equation}
F_{1}(x,k)=\Phi _{0}(x,k)-\frac{\Phi _{0}(x,i\kappa _{1})}{\Delta _{1}(x)}
\int_{-(k_{\Im }-\alpha _{1})\infty }^{x_{1}}\hspace{-20pt}
dx_{1}^{\prime }\ \Phi_{0}(x_{1}^{\prime },x_{2},k)\Psi _{0}(x_{1}^{\prime },
x_{2},i\alpha _{1}).
\label{218}
\end{equation}
Eq. (\ref{218}) shows that $F_{1}$ has the same analytical properties of 
$\Phi _{0}$ and an additional pole for $k=i\alpha _{1}$. In fact, since 
$\Phi_{0}$ is continuous for $k=i\alpha _{1}$, it follows from 
(\ref{xortog}) that 
\begin{equation}
F_{1}(x,k_{\Re }+i(\alpha _{1}+0))-F_{1}(x,k_{\Re }+i(\alpha _{1}-0))
=-2\pi \frac{\Phi _{0}(x,i\kappa _{1})}{\Delta _{1}(x)}\delta (k_{\Re })
\label{233}
\end{equation}
and then 
\[
F_{1}(x,k)=\frac{1}{ik+\alpha _{1}}\frac{\Phi _{0}(x,i\kappa _{1})}{\Delta
_{1}(x)}+\mathrm{reg}\label{F1pole}
\]
where \textrm{reg }denotes terms which are regular in the limit 
$k\rightarrow i\alpha _{1}$. Moreover one can check that there exist limits
\begin{equation}
\lim_{x_{1}\rightarrow \pm \infty
}e^{ikx_{1}+k^{2}x_{2}}F_{1}(x,k)=A_{1}(\pm ,k)  \label{219}
\end{equation}
where 
\begin{equation}
A_{1}(\pm ,k)=1+\frac{(\kappa _{1}-\alpha _{1})}{(ik+\alpha _{1})}\theta
\left( \pm (\kappa _{1}-\alpha _{1})\right)  \label{A1}
\end{equation}
and so the solution of the modified integral equation (\ref{12}) with
potential $u_{1}$ is given by 
\begin{equation}
\chi _{1}(x,k)=e^{^{ikx_{1}+k^{2}x_{2}}}\Phi _{1}(x,k)  \label{Jost}
\end{equation}
with
\begin{equation}
\Phi _{1}(x,k)=\frac{F_{1}(x,k)}{A_{1}(-\mathop{\rm sign}k_{\Im },k)}.
\label{Jost1}
\end{equation}
Indeed, since $\Phi _{1}$ satisfies the heat conduction equation with 
potential $u_{1}$ we have 
\begin{eqnarray*}
&&\int dx^{\prime }
\left( \partial _{x_{1}}G_{0}(x-x^{\prime },k)\right) \ u_{1}(x^{\prime
})\ \chi _{1}(x^{\prime },k)=\int dx^{\prime }\left( \partial
_{x_{1}}G_{0}(x-x^{\prime },k)\right) \times \\
&&\times \left( -\partial _{x_{2}^{\prime }}+
\partial _{x_{1}^{\prime}}^{2}-2ik\partial _{x_{1}^{\prime }}\right)\chi _{1}
(x^{\prime },k)
\end{eqnarray*}
and $\partial _{x_{1}}$ cancels all slowly decaying terms 
of the Green's function (see (\ref{asympx1}) and (\ref{asympx2}))
so we can integrate by parts in the right hand side getting 
\[
\int dx^{\prime }\partial _{x_{1}}G_{0}(x-x^{\prime },k)\ u_{1}(x^{\prime
})\ \chi _{1}(x^{\prime },k)=\partial _{x_{1}}\chi _{1}(x,k). 
\]
It follows that
\begin{eqnarray*}
&&1+\int_{-k_{\Im }\infty }^{x_{1}}dy_{1}\int dx^{\prime }\partial
_{y_{1}}G_{0}(y_{1}-x_{1}^{\prime },x_{2}-x_{2}^{\prime },k)\
u_{1}(x^{\prime })\ \chi _{1}(x^{\prime },k)= \\
&=&1+\int_{-k_{\Im }\infty }^{x_{1}}dy_{1}\partial _{y_{1}}\chi
_{1}(y_{1},x_{2},k)=1+\chi _{1}(x,k)-\lim_{x_{1}\rightarrow -k_{\Im }\infty
}\chi _{1}(x,k)
\end{eqnarray*}
and this completes the proof since, due to definition (\ref{Jost}), 
$
\lim_{x_{1}\rightarrow -k_{\Im }\infty }\chi _{1}(x,k)=1. 
$
Note that from (\ref{A1}) and (\ref{Jost1}) 
and taking into account (\ref{F1pole}), we see that if $\kappa _{1}\alpha
_{1}<0$ $\Phi _{1}$ has no poles in the complex $k$-plane; if $\kappa
_{1}\alpha _{1}>0$ and $0<\alpha _{1}<\kappa _{1}$ or $\kappa _{1}<\alpha
_{1}<0$ it has a pole at $k=i\alpha _{1}$ while if $\alpha _{1}<\kappa
_{1}<0 $ or $0<\kappa _{1}<\alpha _{1}$ it has a pole for $k=i\kappa _{1}$
(in any case the pole corresponds to the one which is smaller by modulo).
So, strictly speaking, $\Phi_{1}$ cannot be considered a Jost solution since
it may have a pole. We will call it Jost-like solution. 

As far as the dual solutions are concerned, in the same way as before we get 
\begin{equation}
\psi _{1}(x,k)=\Psi _{0}(x,k)-\frac{\Psi _{0}(x,i\alpha _{1})}
{\Delta _{1}(x)}\left[ D_{1}(k)+\int_{(k_{\Im }-\kappa _{1})\infty} ^{x_{1}}
\hspace{-20pt}dx_{1}^{\prime}\ \Psi _{0}(x_{1}^{\prime },x_{2},k)\Phi _{0}
(x_{1}^{\prime },x_{2},i\kappa_{1})\right]   \label{2210}
\end{equation}
\begin{equation}
\Upsilon _{1}(x,k)=\Psi _{0}(x,k) -\frac{\Psi _{0}(x,i\alpha _{1})}{\Delta
_{1}(x)}\int_{(k_{\Im }-\kappa _{1})\infty }^{x_{1}}\hspace{-20pt}
dx_{1}^{\prime }\ \Psi_{0}(x_{1}^{\prime },x_{2},k)\Phi _{0}(x_{1}^{\prime },
x_{2},i\kappa _{1}).\label{G1}
\end{equation} 
Note that $\Upsilon _{1}$, as a function
of the spectral parameter $k$, has the same analytical properties of $\Psi
_{0}$ and an additional pole for $k=i\kappa _{1}$. Indeed, using (\ref
{xortog}) one can easily check from (\ref{G1}) that 
\begin{equation}
\Upsilon _{1}(x,k)=\frac{1}{ik+\kappa _{1}}\frac{\Psi _{0}(x,i\alpha _{1})}
{\Delta _{1}(x)}+\mathrm{reg.}  \label{G1pole}
\end{equation}
If we compute its asymptotic behavior, we see that 
\begin{equation}
\lim_{x_{1}\rightarrow \pm \infty }e^{-ikx_{1}-k^{2}x_{2}}\Upsilon
_{1}(x,k)=B_{1}(\pm ,k)
\end{equation}
with 
\begin{equation}
B_{1}(\pm ,k)=1+\frac{(\alpha _{1}-\kappa _{1})}{(ik+\kappa _{1})}\theta
(\pm (\kappa _{1}-\alpha _{1}))  \label{B1}
\end{equation}
and then, to obtain the dual Jost solution, satisfying the modified integral
equation (\ref{13}), we have to consider 
\begin{equation}
\Psi _{1}(x,k)=\frac{\Upsilon _{1}(x,k)}{B_{1}(\mathop{\rm sign}k_{\Im },k)}.
\end{equation}
Using (\ref{B1}) and (\ref{G1pole}) we can prove that $\Psi _{1}$ has no
poles iff $\alpha _{1}\kappa _{1}<0.$ For $0<\alpha _{1}<\kappa _{1}$ or 
$\kappa _{1}<\alpha _{1}<0$ it has a pole at $k=i\alpha _{1}$ while if 
$\alpha _{1}<\kappa _{1}<0$ or $0<\kappa _{1}<\alpha _{1}$ it has a pole
behavior at $k=i\kappa _{1}.$

\section{Two solitons on a generic background\label{2step}}

Now we are going to iterate once the procedure illustrated above. We
consider 
\begin{equation}
\phi _{1}(x,k)=\phi _{0}(x,k)-a_{1}(x)\left[ C_{1}(k)+\int_{-(k_{\Im
}-\alpha _{1})\infty }^{x_{1}}dx_{1}^{\prime }\phi _{0}(x_{1}^{\prime
},x_{2},k)\psi _{0}(x_{1}^{\prime },x_{2},i\alpha _{1})\right]  \label{330}
\end{equation}
and 
\begin{equation}
\psi _{1}(x,k)=\psi _{0}(x,k)-b_{1}(x)\left[ D_{1}(k)+\int_{(k_{\Im }-\kappa
_{1})\infty }^{x_{1}}dx_{1}^{\prime }\psi _{0}(x_{1}^{\prime },x_{2},k)\phi
_{0}(x_{1}^{\prime },x_{2},i\kappa _{1})\right]  \label{350}
\end{equation}
where $\phi _{0}$ and $\psi _{0}$ are, respectively, the Jost solution of
the original spectral problem and of its dual, that is 
\[
\phi _{0}(x,k)=\Phi _{0}(x,k),\qquad \psi _{0}(x,k)=\Psi _{0}(x,k), 
\]
and we introduced 
\begin{eqnarray}
a_{1}(x) &=& \frac{\phi _{0}(x,i\kappa _{1})}{\Delta _{1}(x)}  \qquad 
b_{1}(x) =\frac{\psi _{0}(x,i\alpha _{1})}{\Delta _{1}(x)}  \label{320} \\
\Delta _{1}(x) &=&c_{1}+\int_{-(\kappa _{1}-\alpha _{1})\infty
}dx_{1}^{\prime }\phi _{0}(x_{1}^{\prime },x_{2},i\kappa _{1})\psi
_{0}(x_{1}^{\prime },x_{2},i\alpha _{1})
\end{eqnarray}
and $C_{1}(k)$ and $D_{1}(k)$ are arbitrary functions of $k$ such that 
$C_{1}(ik_{\Im }),\ D_{1}(ik_{\Im })$ are real, so that $\phi _{1}(x,ik_{\Im
})$ and $\psi _{1}(x,ik_{\Im })$ are real and can be used to generate the
binary BT.

We proved that $a_{1}(x)$ and $b_{1}(x)$ are solutions of the spectral
problem (\ref{heatop}) with potential $u_{1}$ and of its dual, respectively.
Now we consider, starting from potential $u_{1}$, a binary BT to generate 
\begin{equation}
u_{2}(x)=u_{1}(x)-2\partial _{x_{1}}^{2}\log \Delta _{2}(x)  \label{u2}
\end{equation}
or 
\begin{equation}
u_{2}(x)=u_{0}(x)-2\partial _{x_{1}}^{2}\log \Delta _{1}(x)\Delta _{2}(x)
\label{3112}
\end{equation}
with 
\begin{equation}
\Delta _{2}(x)=c_{2}+\int_{-(\kappa _{2}-\alpha _{2})\infty
}^{x_{1}}dx_{1}^{\prime }\phi _{1}(x_{1}^{\prime },x_{2},i\kappa _{2})\psi
_{1}(x_{1}^{\prime },x_{2},i\alpha _{2}),  \label{390}
\end{equation}
provided it is possible to choose $\kappa _{2}$ and $\alpha _{2}$ such that
the integral is convergent. 
In the general case, when both $C_{1}(i\kappa_{2})$ and $D_{1}(i\alpha_{2})$
are different from zero, it can be shown 
from (\ref{330}) and (\ref{350}) that the integral in (\ref{390}) is 
convergent if and only if $\kappa _{2}$ and $\alpha _{2}$
are such that the intervals $(\alpha _{1},\kappa _{1})$ and $(\alpha
_{2},\kappa _{2})$ have non void intersection. 
When either one or both are equal to zero less stringent conditions are
required. 
Let us observe that, due to (\ref{217})--(\ref{218})  and 
(\ref{2210})--(\ref{G1}), we have  
\begin{eqnarray*}
\lim_{x_{1}\rightarrow -(\kappa _{2}-\alpha _{2})\infty}\Delta _{2}(x)&=&
c_{2} \\
\lim_{x_{1}\rightarrow (\kappa _{2}-\alpha _{2})\infty}
\Delta _{2}(x)e^{-(\kappa _{2}-\alpha _{2})x_{1}-(\kappa _{2}^{2}-\alpha
_{2}^{2})x_{2}}&=&\frac{A_{1}(\mathop{\rm sign}(\kappa _{2}-\alpha _{2}),
i\kappa _{2})B_{1}(\mathop{\rm sign}(\kappa _{2}-\alpha _{2}),i\alpha _{2})}
{\kappa _{2}-\alpha_{2}} \label{Delta2inf}
\end{eqnarray*}
where the coefficient is given by (\ref{A1})--(\ref{B1}), that is
\begin{eqnarray}
&&A_{1}(\mathop{\rm sign}(\kappa _{2}-\alpha _{2}),i\kappa _{2})B_{1}
(\mathop{\rm sign}(\kappa _{2}-\alpha _{2}),i\alpha _{2})=  \label{A1B1} \\
&=&1+\frac{(\kappa _{1}-\alpha _{1})(\kappa _{2}-\alpha _{2})}{(\alpha
_{1}-\kappa _{2})(\kappa _{2}-\alpha _{1})}\theta ((\kappa _{2}-\alpha
_{2})(\kappa _{1}-\alpha _{1})).  \nonumber
\end{eqnarray}
When $\kappa _{2}$ and $\alpha _{2}$ are both inside the interval of $\kappa
_{1}$ and $\alpha _{1}$ or both outside, the asymptotics (\ref{Delta2inf})
have the same sign at both infinities iff 
\begin{equation}
(\kappa _{2}-\alpha _{2})c_{2}\ge 0,  \label{12e}
\end{equation}
while for one inside and the other one outside the interval $(\alpha
_{1},\kappa _{1})$ a necessary condition for regularity becomes 
\begin{equation}
(\alpha _{1}-\kappa _{1})c_{2}\ge 0.
\end{equation}

\subsection{Properties of potential}

The easiest way to study properties of potential (\ref{3112}) is to solve
the recursion relation and express $\Delta _{1}\Delta _{2}$ in terms of the
Jost solutions of the original spectral problem. Inserting (\ref{330}) and 
(\ref{350}) into (\ref{390}) and integrating by parts using the identity 
\begin{equation}
\frac{\phi _{0}(x,i\kappa _{1})\psi _{0}(x,i\alpha _{1})}{\Delta _{1}(x)}
=-\partial _{x_{1}}\left( \frac{1}{\Delta _{1}(x)}\right) ,  \label{parts}
\end{equation}
and the following condition
\begin{equation}
(\kappa _{2}-\alpha _{2})(\kappa _{1}-\alpha _{1})<0,  \label{altern}
\end{equation}
we can write $\Delta _{1}\Delta _{2}$ as
\begin{equation}
\Delta _{1}(x)\Delta _{2}(x)=\det A_{2}(x)  \label{4231}
\end{equation}
where we introduced the $2\times 2$ matrices $A_{2},$\ $B_{2},$ and$\ C_{2}$
given by 
\begin{equation}
A_{2}(x)=B_{2}(x)+C_{2}  \label{A2(x)}
\end{equation}
and $B_{2}$ and $C_{2}$ with entries
\begin{equation}
c_{11} =c_{1},\qquad c_{22}=c_{2},  \qquad 
c_{12} =C_{1}(i\kappa _{2}),\qquad c_{21}=D_{1}(i\alpha _{2})  \label{c21}
\end{equation}
\begin{equation}
\left( B_{2}\right) _{jl}(x) =\int_{-(\kappa _{l}-\alpha _{j})\infty
}^{x_{1}}dx_{1}^{\prime }\phi _{0}(x_{1}^{\prime },x_{2},i\kappa _{l})\psi
_{0}(x_{1}^{\prime },x_{2},i\alpha _{j}).  \label{503}
\end{equation}
Consequently, from (\ref{3112}) and (\ref{4231}) we obtain for potential 
$u_{2}$ the following expression 
\begin{equation}
u_{2}=u_{0}-2\partial _{x_{1}}^{2}\log \det A_{2}(x).  \label{u2sol}
\end{equation}
Due to the result of Prop.~2, it follows from (\ref{4231}) that when 
$\kappa _{2}$ and $\alpha _{2}$
obey (\ref{altern}) and are such that the intervals $\kappa _{2},\alpha _{2}$
and $\kappa _{1},\alpha _{1}$ are contained one into another, the following
conditions
\begin{equation}
c_{ij}(\kappa _{j}-\alpha _{i})\ge 0,\qquad i,j=1,2,  \label{cond-c}
\end{equation}
ensure that $\Delta _{1}\Delta _{2}$ has no zeros in the $x$-plane and,
consequently, that the dressed potential (\ref{u2sol}) is regular. When the
two intervals, though having non void intersection, are not contained one
into another, these conditions are not sufficient to conclude as before that 
$\Delta _{1}\Delta _{2}$ has no zeros. Then we'll choose the parameters
ordered either according to 
\begin{eqnarray}
\kappa _{1}<\alpha _{2}< \kappa _{2}<\alpha _{1} \nonumber \\
\alpha _{1} <\kappa _{2}<\alpha _{2}<\kappa _{1} 
\label{dentro}
\end{eqnarray}
or to 
\begin{eqnarray}
\kappa _{2} <\alpha _{1}<\kappa _{1}<\alpha _{2}  \nonumber \\
\alpha _{2} <\kappa _{1}<\alpha _{1}<\kappa _{2} .  
\label{fuori}
\end{eqnarray}
Now let us compute asymptotic behavior of potential $u_{2}$ along a generic
direction $d=x_{1}+hx_{2}$. From (\ref{4231}) and using asymptotics of Jost
solutions we get that for $x_{2}\rightarrow \pm \infty $ at fixed $d$ 
\begin{eqnarray}
&&\left( \Delta _{1}\Delta _{2}\right) (d-hx_{2},x_{2})\sim \det C_{2}+ 
\nonumber \\
&&+\frac{c_{22}}{\kappa _{1}-\alpha _{1}}e^{(\kappa _{1}-\alpha
_{1})d+(\kappa _{1}-\alpha _{1})(\kappa _{1}+\alpha _{1}-h)x_{2}}+
\frac{c_{11}}{\kappa _{2}-\alpha _{2}}e^{(\kappa _{2}-\alpha _{2})d+(\kappa
_{2}-\alpha _{2})(\kappa _{2}+\alpha _{2}-h)x_{2}}  \nonumber \\
&&+\frac{c_{12}}{\alpha _{2}-\kappa _{1}}e^{(\kappa _{1}-\alpha
_{2})d+(\kappa _{1}-\alpha _{2})(\kappa _{1}+\alpha _{2}-h)x_{2}}+
\frac{c_{21}}{\alpha _{1}-\kappa _{2}}e^{(\kappa _{2}-\alpha _{1})d+(\kappa
_{2}-\alpha _{1})(\kappa _{2}+\alpha _{1}-h)x_{2}}  \nonumber \\
&&+K e^{(\kappa_{1}-\alpha _{1}+\kappa _{2}-\alpha _{2})d
+(\kappa _{1}-\alpha _{1})(\kappa _{1}+\alpha_{1}-h)x_{2}
+(\kappa _{2}-\alpha _{2})(\kappa_{2}+\alpha _{2}-h)x_{2}}  \label{gend}
\end{eqnarray}
where we introduced
\[
K=-(\kappa_{2}-\kappa_{1})(\alpha_{2}-\alpha_{1})\prod_{l,j=1,2}
\frac{1}{\kappa_{j}-\alpha_{l}}.
\]

It is tedious but not difficult to prove that along any direction
$d=x_{1}+hx_{2}$ with $h\ne \kappa_{i}+\alpha_{j}, \  
\kappa_{i}+\kappa_{i+1}, \  \alpha_{i}+\alpha_{i+1}$ 
for $i,j=1,2$ and where $i+1$ is intended modulo 2, the contribution of
$\Delta_{1}\Delta_{2}$ to the potential $u_{2}$ is exponentially decaying.

However, the asymptotic behavior (\ref{gend}) along these ``special'' 
directions depends widely on matrix $C_{2}$, in the sense that not only the 
phase shift but also the number and directions of the solitons emerging from 
the background are determined by the entries of this matrix.
We proved that if $c_{ij}\ne 0$ but $c_{i+1,j}=c_{i,j+1}=c_{i+1,j+1}=0$
the potential is decaying along all directions but the one corresponding
to $h=\alpha_{i}+\kappa_{j}$ and along this direction it exhibits 
asymptotically the expected soliton-like behavior.
For $c_{ii}\ne 0$ and $c_{i,i+1}=0$ or $c_{ii}=0$ and $c_{i,i+1}\ne0$ 
$i=1,2$ we have two ``shifted'' solitons along the directions 
$h=\alpha_{i}+\kappa_{i}$ $i=1,2$ or, respectively, 
$h=\alpha_{i}+\kappa_{i+1}$ $i=1,2$.
When $c_{ii}\ne 0$, $c_{i,i+1}\ne 0$ but $c_{i+1,i+1}=c_{i+1,i}=0$ or
$c_{ii}\ne 0$,  $c_{i+1,i}\ne 0$ but $c_{i+1,i+1}=c_{i,i+1}=0$ matrix
$C_{2}$ is singular and in these cases we have three ``tails'' along
the directions 
$h=\alpha_i+\kappa_i, \ \alpha_i +\kappa_{i+1},\ \kappa_{i}+\kappa_{i+1}$ or,
respectively,
$h=\alpha_i+\kappa_i, \ \alpha_{i+1}+\kappa_{i},\ \alpha_{i}+\alpha_{i+1}$.
When all entries but one, say $c_{ij}$, are different from zero, the soliton
directions are given by $h=\kappa_{i}+\kappa_{i+1},\ \alpha_{i}+\alpha_{i+1}$
and $h=\alpha_{i}+\kappa_{i}, \alpha_{i+1}+\kappa_{i+1}$ if $i\ne j$ and
$h=\alpha_{i+1}+\kappa_{i}, \alpha_{i}+\kappa_{i+1}$ if $i=j$.
Finally, when $c_{ij}\ne 0$ for all $i,j=1,2$ the potential exhibits
asymptotically two ``shifted'' solitons along the directions 
$h=\alpha_{i}+\alpha_{i+1}$ and $h=\kappa_{i}+\kappa_{i+1}$.

\subsection{Darboux transform}

As far as the Darboux transforms of $\phi _{1}$ and $\psi _{1}$ are
concerned, we can formally define 
\begin{equation}
\phi _{2}(x,k)=\phi _{1}(x,k)-a_{2}(x)\left[ C_{2}(k)+\int_{-(k_{\Im
}-\alpha _{2})\infty }^{x_{1}}\hspace{-5pt}
dx_{1}^{\prime }\phi _{1}(x_{1}^{\prime
},x_{2},k)\psi _{1}(x_{1}^{\prime },x_{2},i\alpha _{2})\right]   \label{505}
\end{equation}
and 
\begin{equation}
\psi _{2}(x,k)=\psi _{1}(x,k)-b_{2}(x)\left[ D_{2}(k)+\int_{(k_{\Im }-\kappa
_{2})\infty }^{x_{1}}\hspace{-5pt}
dx_{1}^{\prime }\psi _{1}(x_{1}^{\prime },x_{2},k)\phi
_{1}(x_{1}^{\prime },x_{2},i\kappa _{2})\right]   \label{506}
\end{equation}
with 
\begin{equation}
a_{2}(x) =\frac{\phi _{1}(x,i\kappa _{2})}{\Delta _{2}(x)}, \qquad 
b_{2}(x) =\frac{\psi _{1}(x,i\alpha _{2})}{\Delta _{2}(x)}.
\label{507}
\end{equation}
However, due to (\ref{A1}) and (\ref{B1}), the integrals in 
(\ref{505})--(\ref{506}) are convergent for arbitrary $k$ only if 
\begin{equation}
(\alpha _{2}-\kappa _{1})(\alpha _{2}-\alpha _{1}) <0  \qquad
(\alpha _{1}-\kappa _{2})(\kappa _{1}-\kappa _{2}) <0,  \label{convDarb2}
\end{equation}
and these conditions require the parameters to be ordered according to (\ref
{dentro}). In other words, like in KPI case \cite{steklov}, the recursion 
procedure for getting the solutions is well defined only for a proper choice 
of order for parameters, even though we proved that the potential is well 
defined and regular in more general situations.

Now we have to solve the recursion procedure. If we substitute into 
(\ref{505}) the expressions (\ref{330}) and (\ref{350}) for $\phi _{1}$ and 
$\psi_{1}$ and integrate by parts using again (\ref{parts}), we find 
\begin{equation}
\phi _{2}(x,k)=\frac{1}{\det A_{2}(x)}\left| 
\begin{array}{lll}
A_{11}(x) & A_{12}(x) & \beta _{1}(x,k)+C_{1}(k) \\ 
A_{21}(x) & A_{22}(x) & \beta _{2}(x,k)+C_{2}(k) \\ 
\phi _{0}(x,i\kappa _{1}) & \phi _{0}(x,i\kappa _{2}) & \phi _{0}(x,k)
\end{array}
\right|   \label{phi2}
\end{equation}
where matrix $A_{2}$ is given by (\ref{A2(x)}) and 
\begin{equation}
\beta _{j}(x,k)=\int_{-(k_{\Im }-\alpha _{j})\infty }^{x_{1}}dx_{1}^{\prime
}\phi _{0}(x_{1}^{\prime },x_{2},k)\psi _{0}(x_{1}^{\prime },x_{2},i\alpha
_{j}).  \label{betaj}
\end{equation}
Clearly, since $\phi _{2}$ given by (\ref{phi2}) is a solution of the
spectral problem for $u_{2}$ for any $C_{1}(k)$ and $C_{2}(k)$,
properties of determinants ensure that also 
\begin{equation}
F_{2}(x,k)=\frac{1}{\det A_{2}(x)}\left| 
\begin{array}{lll}
A_{11}(x) & A_{12}(x) & \beta _{1}(x,k) \\ 
A_{21}(x) & A_{22}(x) & \beta _{2}(x,k) \\ 
\phi _{0}(x,i\kappa _{1}) & \phi _{0}(x,i\kappa _{2}) & \phi _{0}(x,k)
\end{array}
\right|   \label{F2}
\end{equation}
and 
\begin{equation}
f_{2}(x,k)=\frac{1}{\det A_{2}(x)}\left| 
\begin{array}{lll}
A_{11}(x) & A_{12}(x) & C_{1}(k) \\ 
A_{21}(x) & A_{22}(x) & C_{2}(k) \\ 
\phi _{0}(x,i\kappa _{1}) & \phi _{0}(x,i\kappa _{2}) & 0
\end{array}
\right|   \label{f2}
\end{equation}
are solutions of the same spectral problem and 
\begin{equation}
\phi _{2}(x,k)=F_{2}(x,k)+f_{2}(x,k).  \label{F2+f2}
\end{equation}
One can easily obtain the analogous expression for dual solutions, that is 
\begin{equation}
\psi _{2}(x,k)=\frac{1}{\det A_{2}(x)}\left| 
\begin{array}{lll}
A_{11}(x) & A_{12}(x) & \beta _{1}^{d}(x,k)+D_{1}(k) \\ 
A_{21}(x) & A_{22}(x) & \beta _{2}^{d}(x,k)+D_{2}(k) \\ 
\psi _{0}(x,i\alpha _{1}) & \psi _{0}(x,i\alpha _{2}) & \psi _{0}(x,k)
\end{array}
\right|   \label{psi2}
\end{equation}
and 
\begin{equation}
\Upsilon _{2}(x,k)=\frac{1}{\det A_{2}(x)}\left| 
\begin{array}{lll}
A_{11}(x) & A_{12}(x) & \beta _{1}^{d}(x,k) \\ 
A_{21}(x) & A_{22}(x) & \beta _{2}^{d}(x,k) \\ 
\psi _{0}(x,i\alpha _{1}) & \psi _{0}(x,i\alpha _{2}) & \psi _{0}(x,k)
\end{array}
\right|   \label{G2}
\end{equation}
where 
\begin{equation}
\beta _{j}^{d}(x,k)=\int_{(k_{\Im }-\kappa _{j})\infty
}^{x_{1}}dx_{1}^{\prime }\psi _{0}(x_{1}^{\prime },x_{2},k)\phi
_{0}(x_{1}^{\prime },x_{2},i\kappa _{j}).  \label{betad}
\end{equation}
For $F_{2}$ and $\Upsilon _{2}$ one has the expressions 
\begin{equation}
F_{2}(x,k)=F_{1}(x,k)-a_{2}(x)\int_{-(k_{\Im }-\alpha _{2})\infty
}^{x_{1}}dx_{1}^{\prime }F_{1}(x_{1}^{\prime },x_{2},k)\psi
_{1}(x_{1}^{\prime },x_{2},i\alpha _{2}),  \label{517}
\end{equation}
\begin{equation}
\Upsilon _{2}(x,k)=\Upsilon _{1}(x,k)-b_{2}(x)\int_{(k_{\Im }-\kappa
_{2})\infty }^{x_{1}}dx_{1}^{\prime }\Upsilon _{1}(x_{1}^{\prime
},x_{2},k)\phi _{1}(x_{1}^{\prime },x_{2},i\kappa _{2}).  \label{518}
\end{equation}
Then, if we compute 
\begin{equation}
A_{2}(\pm ,k)=\lim_{x_{1}\rightarrow \pm \infty
}e^{ikx_{1}+k^{2}x_{2}}F_{2}(x,k)  \label{A2}
\end{equation}
we find
\[
A_{2}(\pm ,k)=A_{1}(\pm ,k)+\frac{\kappa _{2}-\alpha _{2}}{ik+\alpha _{2}}
A_{1}(\pm ,k)B_{1}(\pm ,i\alpha _{2})\theta (\pm (\kappa _{2}-\alpha _{2}))
\]
where we used that, due to (\ref{dentro}), 
\[
\lim_{x_{1}\rightarrow \pm \infty }\frac{e^{(\kappa _{2}-\alpha _{1})x_{1}}}
{\left( 1+e^{(\kappa _{2}-\alpha _{2})x_{1}}\right) \left( 1+e^{(\kappa
_{1}-\alpha _{1})x_{1}}\right) }=0.
\]
Taking into account (\ref{altern}) and (\ref{B1}), we can write the
following recursion relation 
\begin{equation}
A_{2}(\pm ,k)=A_{1}(\pm ,k)\left[ 1+\frac{\kappa _{2}-\alpha _{2}}{ik+\alpha
_{2}}\theta (\pm (\kappa _{2}-\alpha _{2}))\right] .  \label{recA}
\end{equation}
Analogously, from (\ref{518}) one can check that $B_{2}(\pm ,k)$ defined as 
\begin{equation}
B_{2}(\pm ,k)=\lim_{x_{1}\rightarrow \pm \infty
}e^{-ikx_{1}-k^{2}x_{2}}\Upsilon _{2}(x,k)  \label{B2}
\end{equation}
obeys the following recursive relation 
\begin{equation}
B_{2}(\pm ,k)=B_{1}(\pm ,k)\left[ 1+\frac{\alpha _{2}-\kappa _{2}}{ik+\kappa
_{2}}\theta (\pm (\kappa _{2}-\alpha _{2}))\right] .  \label{recB}
\end{equation}
Then we obtain the Jost solution and the dual Jost solution for potential 
$u_{2}$ as 
\begin{eqnarray*}
\Phi _{2}(x,k) &=&\frac{F_{2}(x,k)}{A_{2}(-k_{_{\Im }},k)} \\
\Psi _{2}(x,k) &=&\frac{\Upsilon _{2}(x,k)}{B_{2}(k_{_{\Im }},k)},
\end{eqnarray*}
since it is easy to check that they satisfy integral equations (\ref{12})
and (\ref{13}) with $u$ substituted by $u_{2}$.

\section*{Acknowledgments}
I am very grateful to M.~Boiti, F.~Pempinelli and A.~K.~Pogrebkov for 
help, suggestions and fruitful discussions.


\begin{thebibliography}{99}

\bibitem{Kadomtsev-P}  B.~B.~Kadomtsev and V.~I.~Petviashvili, {\it Sov.
Phys. Doklady } {\bf 192 }, 539 (1970).

\bibitem{dryuma74}  V.~S.~Dryuma, {\it Sov. Phys. J. Exp. Theor. Phys.
Lett.} {\bf 19}, 381 (1974).

\bibitem{zakharov74}  V.~E.~Zakharov and A.~B.~Shabat, {\it Funct. Anal.
Appl. } {\bf 8}, 226 (1974).

\bibitem{ABYF}  M. Ablowitz, D. Bar Yacoov, A.~S. Fokas, {\it Stud. Appl.
Math.} {\bf 69}, 135 (1983).

\bibitem{ZManakov}  V.~E.~Zakharov, S.~V.~Manakov, {\it Sov. Sci. Rev.
Phys. Rev.} {\bf 1}, 133 (1979).

\bibitem{Manakov}  S.~V.~Manakov, {\it Physica} {\bf D3}, 420 (1981).

\bibitem{FokasAblowitz}  A.~S.~Fokas and M.~J.~Ablowitz, {\it Stud. Appl.
Math.} {\bf 69}, 211 (1983).

\bibitem{Grinevich}  P.~G. Grinevich, S.~P. Novikov, {\it Func. Anal.
Appl.} {\bf 22}, 23 (1988).

\bibitem{nsreshi}  M.~Boiti, F.~Pempinelli, A.~K.~Pogrebkov, and
M.~C.~Polivanov, {\it Theor. Math. Phys.} {\bf 93}, 1200 (1992).

\bibitem{KPtmf}  M.~Boiti, F.~Pempinelli, and A.~Pogrebkov, {\it Theor.
Math. Phys.} {\bf 99}, 511 (1994).

\bibitem{proc95}  M.~Boiti, F.~Pempinelli, and A.~Pogrebkov, in {\em
Nonlinear Physics. Theory and Experiment}, edts. E. Alfinito, M.~Boiti,
L.~Martina, and F.~Pempinelli, World Scientific Pub. Co., Singapore (1996),
pp. 37-52.

\bibitem{KPlett}  M.~Boiti, F.~Pempinelli, and A.~Pogrebkov, {\it
Inverse Problems} {\bf 13}, L7 (1997).

\bibitem{towards}  M.~Boiti, F.~Pempinelli, A.~Pogrebkov, and B. Prinari, 
{\it Theor. Math. Phys.} {\bf 116}, 741 (1998).

\bibitem{steklov}  M. Boiti, F. Pempinelli, A.K. Pogrebkov, B. Prinari, 
{\em Proceedings of the Steklov Institute of Mathematics} {\bf 126},
42-62 (1999).

\bibitem{Clarksonbook}  M.~J.~Ablowitz and P.~A.~Clarkson, 
{\em Solitons, nonlinear evolution equations and inverse scattering}, Lecture
Notes Series {\bf 49}, University of Cambridge, Cambridge (1991).


\bibitem{Wick}  M.V.~Wickerhauser, {\it Commun. Math. Phys.} {\bf 108},
67-89 (1987).

\bibitem{matveev}  V.~B.~Matveev and M.~A.~Salle, {\em Darboux
Transformations and Solitons}, Springer, Berlin (1991).

\bibitem{salle}  M.~A.~Salle, PhD Thesis, Leningrad.

\bibitem{darboux}  M.~Boiti, F.~Pempinelli, A.~K.~Pogrebkov, and
M.~C.~Polivanov, {\it Inverse problems} {\bf 7}, 43 (1991).

\end{thebibliography}
\end{document}